\documentclass[manuscript]{aastex}

\usepackage{natbib}
\shorttitle{Seismic solar radius and irradiance variations}
\shortauthors{Jain et al.}

\begin{document}

\title{	Sun's Seismic Radius as Measured from the Fundamental Modes of Oscillations 
and its Implications for the TSI Variations}

\author{Kiran Jain, S.C. Tripathy and F. Hill}
\affil{National Solar Observatory$^1$, 3665 Discovery Drive, Boulder, CO 80303, USA}
\email{kjain@nso.edu, stripathy@nso.edu, fhill@nso.edu}
\altaffiltext{1}{The National Solar Observatory is operated by the 
Association of Universities for Research in Astronomy under a cooperative
agreement with the National Science Foundation, for the benefit of the 
astronomical community. }

\begin{abstract}
In this letter, we explore the relationship between the solar seismic radius and 
total solar irradiance (TSI) during  last two solar cycles using the uninterrupted data from 
space-borne instruments onboard {\it SoHO} and {\it SDO}. 
The seismic radius is calculated from the fundamental ({\it f}) modes of solar oscillations
utilizing the observations from {\it SoHO}/MDI and {\it SDO}/HMI, and the total solar 
irradiance measurements are obtained from {\it SoHO}/VIRGO.
 Our study suggests that the  major contribution to the  TSI variation
arises from the changes in magnetic field while the radius variation plays a secondary role. 
 We find that the solar irradiance increases with decreasing seismic radius,
however the anti-correlation between them is moderately weak. 
The estimated maximum change in seismic radius during a solar cycle is about 5 kilometers,
and  is consistent in both solar cycles 23 and 24.
Previous studies suggest a radius change at the surface of the order of 0.06 arcsecond
to explain the 0.1\% variation in the TSI values during the solar cycle, however
our inferred seismic radius change is significantly smaller, hence the
TSI variations can not be  fully explained by the temporal changes in seismic radius.

\end{abstract}


\section{Introduction}

The solar radius is one of the most fundamental parameters for the precise understanding of Sun's properties.
 Its accurate measurement is important for determining the Sun's composition, structure
 as well as  the rotation rate \citep{Bahcall01}.  Radius measurements at different wavelengths further enable a better
understanding of the solar atmosphere \citep{Thuillier17, Menezes17}. Several authors have investigated the 
possible relationship between the variability of solar radius and the total energy output or total 
solar irradiance (TSI) \citep{Sofia79,Frohlich84,Pap01}. It is well known that the
ultimate source of solar energy is nuclear reactions taking place in the center of the Sun and the
rate of these reactions is almost constant on the time scales of millions of years. On
the other hand,  TSI measurements from space clearly show variability on a time scale of minutes 
to a 11-year solar cycle, thus
there would be some intermediate factors that are responsible for both relatively shorter and longer time-scale.
It is suggested that the solar radius variation could be one
of the factors that might be responsible for the TSI variation, thus the time-dependent 
radius measurements  should be considered while modeling the TSI variability.  In
order to quantify the relation between them, \citet{Sofia98}
argued that 0.1\% change in TSI could be explained by the radius change of 0.06 arcsecond.

In general, two terms are in use for the solar radius;
 the {\it physical} or {\it true} radius and the {\it seismic} or {\it acoustic} radius.  While the former has
records dating back to 18th century \citep[][and references therein]{Vaquero16},  the concept of seismic radius is only a 
 couple of decades old \citep{Schou97}. 
 The measurements of physical  radius are primarily made 
using  solar limb  measurements during planetary transits or the solar disk occultation; however 
 its precise value is still a matter of debate with differences of several tenths of an arcsecond 
  \citep[e.g.,][and references therein]{Emilio15,Rozelot15}. These differences are mainly 
attributed to the type of measurement techniques and the instrument calibration.
Similar to physical radius, the seismic radius measurements also vary marginally with the analysis method.

 The seismic radius is determined by calibrating the radius in a solar model to match the observed 
frequencies. Hence,  it gives the value at a layer that defines  the surface in the Solar model.
Several authors used solar models to estimate the 
 true radius and obtained much smaller values \citep{Brown98, Antia98, Tripathy99}.
 In addition, the near-surface uncertainties in the solar models also contribute to different values.
Analytically, it is estimated from the solar oscillation data by applying different methods. 
In most studies, the global frequencies
of surface-gravity or fundamental ({\it f}) modes are used to quantify the values of seismic radius 
\citep{Schou97,  Antia00, Dziembowski01, Antia04,Dziembowski05}. 
 It was suggested that the {\it f}-mode frequencies are modified by the changes
in both magnetic field and the solar seismic radius.
Since the change in  seismic radius is estimated assuming that the fractional change in 
radius is uniform in the range of sensitivity of the method, its value corresponds to 
 the change at a radius where the {\it f} modes are concentrated.
Although  most of the studies were mainly aimed at determining the precise value of
the solar radius, the obtained values were significantly smaller than the true radius.
 While abovementioned studies primarily utilized modes in the intermediate-degree range, \citet{Kholikov08} analyzed
spherical harmonic coefficient time series of low-degree {\it p} modes in the  range $\ell$ = 0 -- 3 
and calculated autocorrelation function  to infer the acoustic radius.  
 \citet{Irene09} exploited the method of acoustic holography and 
 analyzed the propagation of wave packets to infer the variation 
in seismic radius in cycle 23 by applying the technique of acoustic holography. 
 Although these studies provide different quantative estimates of  change in the seismic radius with time,
 they converge to a single conclusion that  its variation is anti correlated with the phase of the solar cycle.
Moreover, some of these studies  exclude the effect of strong magnetic field in the  determination of seismic
radius and hence display a strong negative correlation between seismic
radius and the solar activity.

Our aim in this paper is to study the temporal variation in seismic radius during last two solar cycles 
 as determined by the  {\it f}-mode frequencies. 
 It is  well accepted that the seismic radius measurements obtained from  {\it f} modes
do not represent the true solar radius but provide its value at a depth of several megameters
below the photosphere.  Here  we discuss the  relative variation of the seismic radius instead 
of its true value. We must emphasize that these variations
represent the changes in the thermal structure of upper convection zone only.
We further compare the seismic radius changes  with the 
variation in TSI. The paper is organized as follows: we briefly describe data and the
method in Section 2. The results are discussed in Section 3 followed by a summary in Section 4.

\section{Data and Analysis}

\subsection{Time series of mode frequencies }

We use  {\it f}-mode frequencies in the spherical harmonic degree $\ell \le$ 300 from 
 the medium-$\ell$ program of {\it Solar and Heliospheric Observatory (SoHO)}/Michelson Doppler Imager
 \citep[MDI;][]{mdi}  and {\it Solar Dynamics Observatory 
(SDO)}/Helioseismic and Magnetic Imager  \citep[HMI;][]{hmi}, covering a period of
about 21 years, i.e. nearly two solar cycles (23~--~24), starting from mid 1996 to mid 2017.  
In total, 74 MDI (May 1, 1996~--~April 24, 2011) and 36 HMI (April 30, 2010~--~June 3, 2017) data sets 
are used. Note that there are 5 overlapping sets between MDI and HMI
from April 30, 2010 to April 24, 2011. Each data set is produced from 72 day long time series and 
the frequency table consists of 
centroid frequencies $\nu_{n,\ell}$ and splitting coefficients $a_i$ where
 each ${n,\ell}$ multiplet is represented by a polynomial expansion
\begin{equation}
\nu_{n,\ell,m} = \nu_{n,\ell} + \sum_{i=1}^{i_{max}} a_i(n,\ell)
P_i^{(\ell)}(m),
  \label{eq2}
\end{equation}
where $P_i^{(\ell)}(m)$'s are orthogonal polynomials of degree $i$  and
${i_{max}}$ is the number of $a$ coefficients used in determining
frequencies. In this paper, we use frequency tables for ${i_{max}}$ = 18.  The remaining symbols in equation (1) have 
their usual meanings. Note that the frequencies from both 
the instruments were calculated by using the same approach, hence these are not biased by any computational method.
Since we are interested in {\it f}-mode frequencies, we analyze $n$ = 0 modes only.

It is worth mentioning that the historic fits to the MDI data suffered from an artificially introduced 
 1-year periodicity due to the orbital period of the Earth  \citep{Schou02}. In
 this work, we use  improved frequency data from MDI \citep{Larson15, Larson18}, which include improvements in a number 
 of geometric corrections made during 
spherical harmonic decomposition; updated routines for generating window functions, detrending time series, 
filling gaps; horizontal displacement at the solar surface, and distortion of eigenfunctions by
differential rotation \citep{Larson15}. As a result, the periodicity in the historic MDI 
{\it f}-mode frequencies \citep[e.g., discussed in][]{Jain03} has
been attenuated significantly in the improved  frequencies, which are used in this paper.

\subsubsection{Combining time series from {\it SoHO}/MDI and {\it SDO}/HMI}

Since both MDI and HMI do not cover the period of two solar cycles independently, 
one of the major tasks in this study is to combine frequency data
from two missions, which use different spectral lines in the solar photosphere.   The 
MDI  observations are in the \ion{Ni}{1} 6768 \AA~line as opposed to HMI observations 
in \ion{Fe}{1} 6173 \AA~line;  the formation height of  \ion{Fe}{1} line is lower in atmosphere 
than the \ion{Ni}{1} line. While mode amplitudes tend to decrease with increasing height, 
the oscillation frequencies and  life-time of the modes are independent of the spectral 
line used in observations \citep{Jain06a}. Although the same peak-bagging method is used to 
calculate the frequencies from both missions, different instruments may also introduce some 
instrument-related bias. Thus, it is important to examine the differences between mode 
frequencies from both missions. For this purpose, we use 5 overlapping 72-d sets from MDI and 
HMI covering the observation for about a year, i.e.,  from mid-2010 to mid-2011. Each set is
represented by a unique data set identifier number,  which corresponds to the day number 
relative to the MDI epoch of 1993 Jan 1 00:00:00\_TAI.

\begin{figure}   
   \centerline{
\includegraphics[scale=0.7]{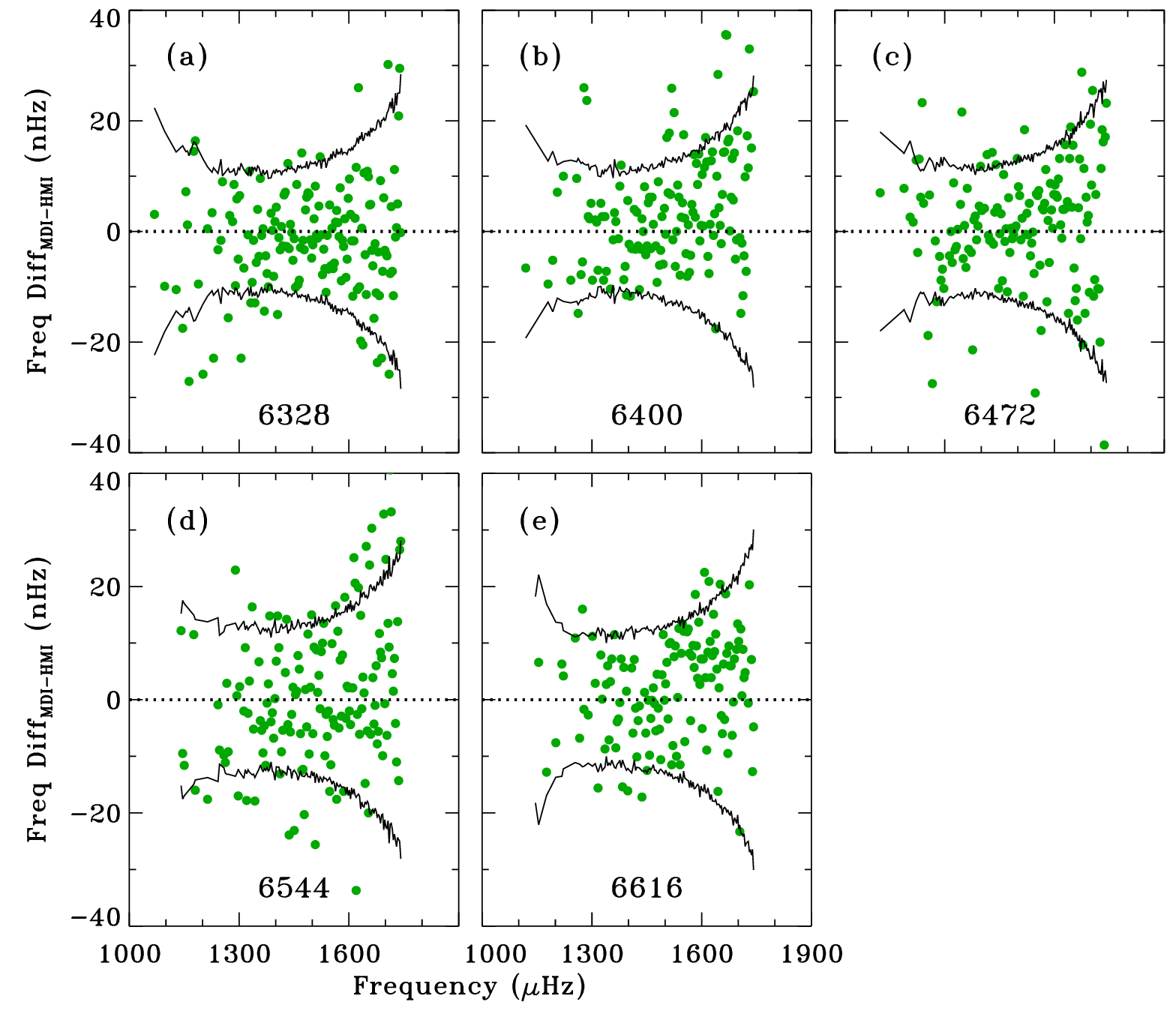}
}
   \centerline{
\includegraphics[scale=0.7]{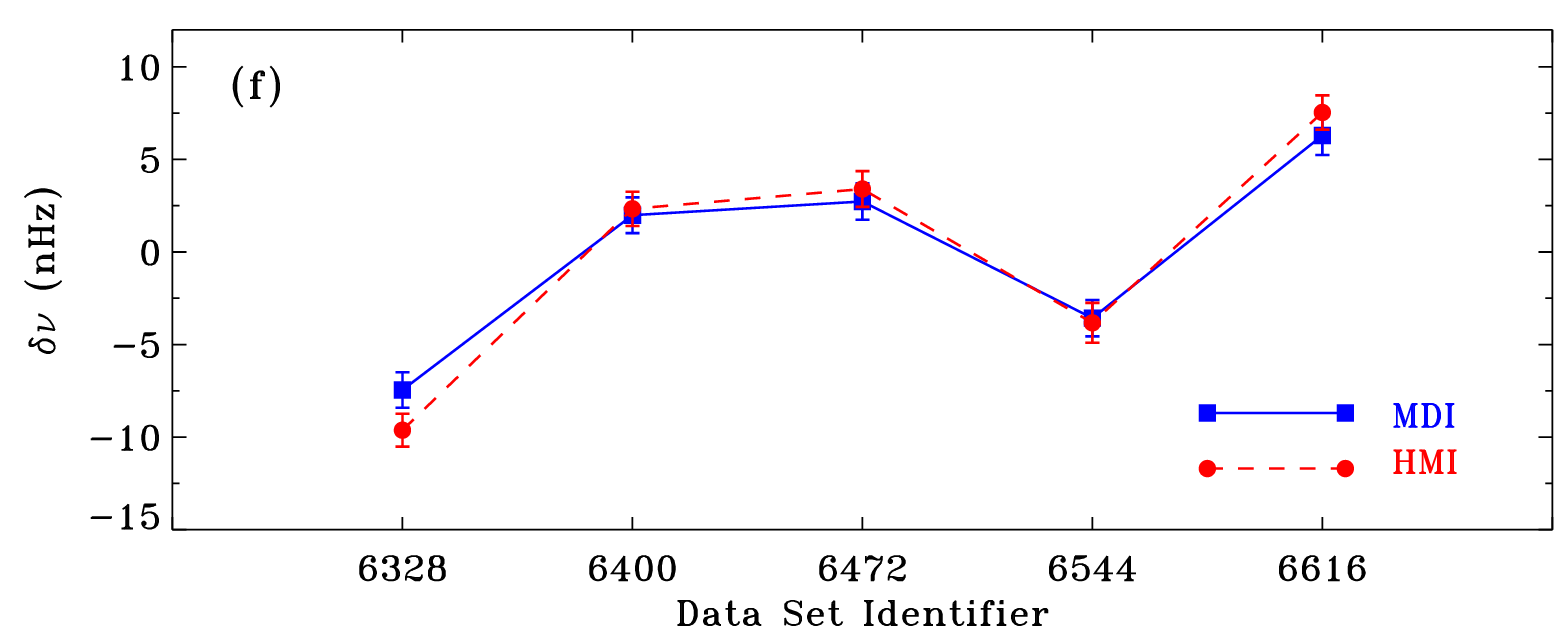}
}
            \caption{(a -- e) Symbols showing difference in {\it f}-mode frequencies for the
 simultaneous observations of MDI and HMI. Each dataset is represented by a unique identifier (shown in each figure),
 which corresponds to the
 day number relative to MDI epoch of 1993 Jan 1 00:00:00\_TAI. Solid lines represent the mean error. (f) Total 
frequency shift calculated for {\it f} modes present in 5 overlapping MDI and HMI data sets. The reference
frequencies are defined as the frequencies averaged over all 5 datasets for individual instrument. 
}
   \label{common}
   \end{figure}

The frequency differences for individual modes in all five sets are shown in Figure~\ref{common} 
(a~--~e). In each case, it is seen that the difference in frequencies (shown by symbols) for most 
modes are less than the mean error (shown by solid lines). The mean frequency shifts with reference 
to the average frequency of  112 modes available in all 10 data sets are plotted in Figure~\ref{common}(f). 
It is evident that there are small differences in the mean shifts for individual epochs, however 
majority of these lie within 1$\sigma$ error. This close agreement between frequencies from simultaneous
observations from both missions allows us to combine two data series to form an uninterrupted long 
data series for about two solar cycles. Since duty cycles for the gap-filled timeseries 
of HMI were higher than the MDI, we prefer to use HMI frequencies for the overlapping period.  Hence 
the MDI frequencies for the period 1996 May 1~--~2010 April 29 and the HMI frequencies from 2010 
April 30 to 2017 June 3 are used in this study.

\subsection{Seismic solar radius and the TSI time series}

\begin{figure}   
   \centerline{
\includegraphics[scale=0.75]{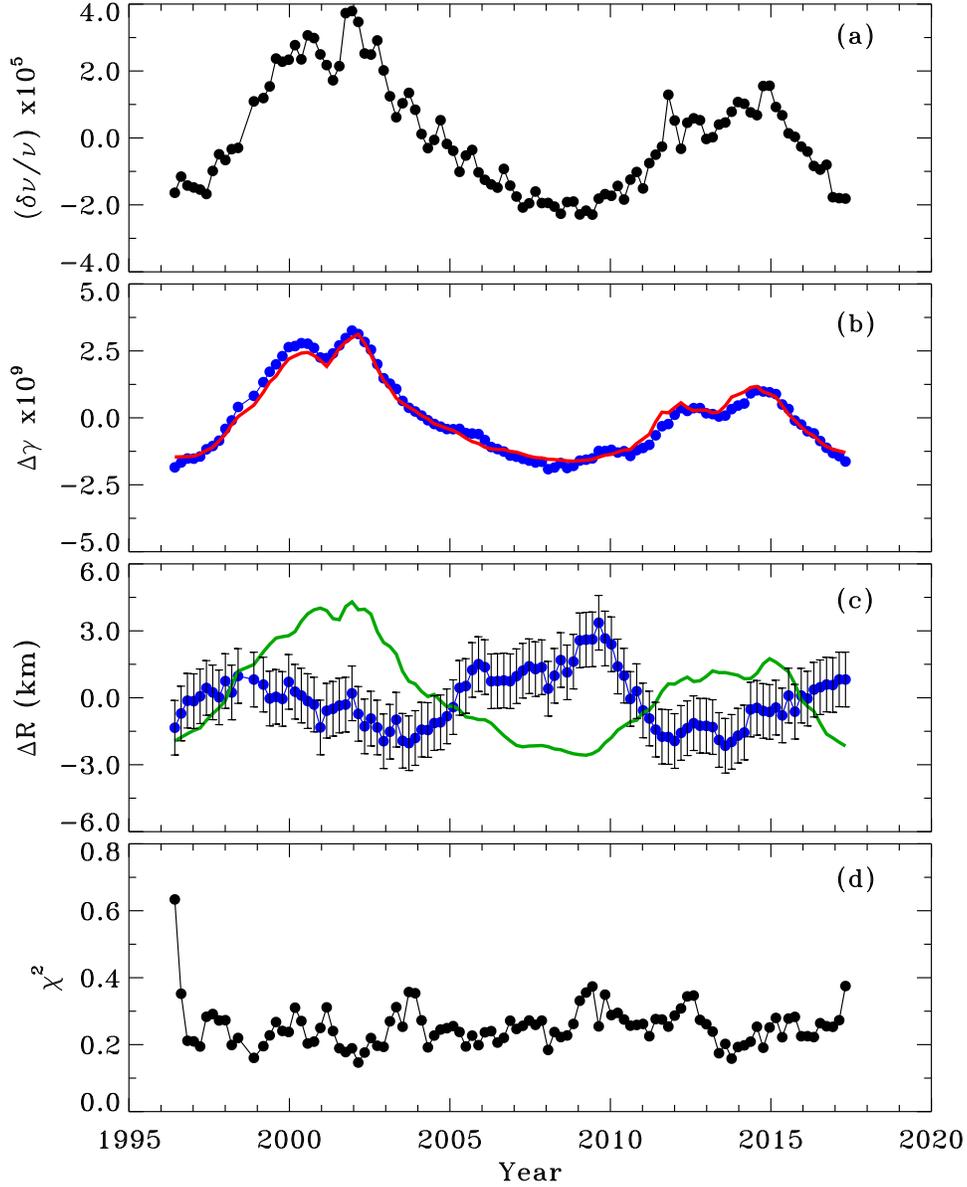}
}
            \caption{Temporal variation  of (a) the  average fractional frequency shifts, 
            (b) the surface term, $\Delta \gamma$, (c) solar seismic radius, $\Delta R$ 
             calculated from the {\it f}-mode frequencies, and (d) the $\chi^2$ per degree of freedom. Blue symbols  in panels (b) and (c)  
             represent the calculated values while red solid line in panel (b) is the smoothed variation of 
            scaled 10.7 cm radio flux and green line in panel c is for the smoothed TSI values. 
            Errors in panels a and b are smaller than the size of the symbols and those in panel (c)
            are the standard deviation in estimated $\Delta R$ values.
}
   \label{radius}
   \end{figure}

The variation in seismic radius with time is estimated from the following relation 
\citep{Dziembowski01},
\begin{equation}
\Delta\nu_{\ell} =- \frac{3}{2} \frac{\Delta R}{R_{Sun}}\nu_{\ell} +\frac{\Delta\gamma}{I_{\ell}}
  \label{eq1}
\end{equation}
where $\Delta R$ is the change in seismic radius inferred from a set of $f$ modes and $\Delta\gamma$ measures
the contribution from surface term.   The first term on the right represents the radius contribution 
 ($\Delta\nu_{R}$) and the second term is surface contribution ($\Delta\nu_\gamma$).
 The mode inertia, $I_{\ell}$, used here is taken from
the standard solar model `BS05'  \citep{BS05}. The values of $\Delta R$ and  $\Delta\gamma$ 
for each epoch are obtained using the least-square method applied to  Equation~\ref{eq1}. 
 Since each mode is trapped at a different layer \citep[e.g.,][]{Sofia05},
the estimation of seismic radius depends on the choice of  modes. Thus, 
 we use 72 {\it f} modes in the $\ell$  range of 216 to 299 which are present in all epochs. This
criterion of selecting modes is important to infer the  true variation in mean seismic radius
 in all data sets.  The temporal variation of calculated average fractional 
frequency shifts, $\delta\nu/\nu$, is plotted in Figure~\ref{radius}(a).   It is evident
 that a 1-year periodicity still exists in frequency timeseries, hence it is removed
 from the timeseries of each mode before fitting  Equation~\ref{eq1} for each epoch. The 
 estimated $\Delta R$, $\Delta\gamma$  and the $\chi^2$ per degree of freedom in Figures~\ref{radius}(b), 
 \ref{radius}(c) and \ref{radius}(d), respectively.
  The $\chi^2$ values fluctuate around 0.2  for all epochs (except near edges due to smoothing)  indicating
 that there are some uncertainties involved in the fitting of Equation~\ref{eq1} and these
 are comparable for both instruments.    We have also shown the scaled variations of
10.7 cm radio flux \citep[$F_{10.7}$;][]{Tapping13}, a proxy for magnetic field, and the TSI  from 
{\it SoHO}/VIRGO \citep{Frohlich97} 
in panels (b) and (c), respectively.

\section{Results and Discussion}

\begin{figure}   
   \centerline{
\includegraphics[scale=0.8]{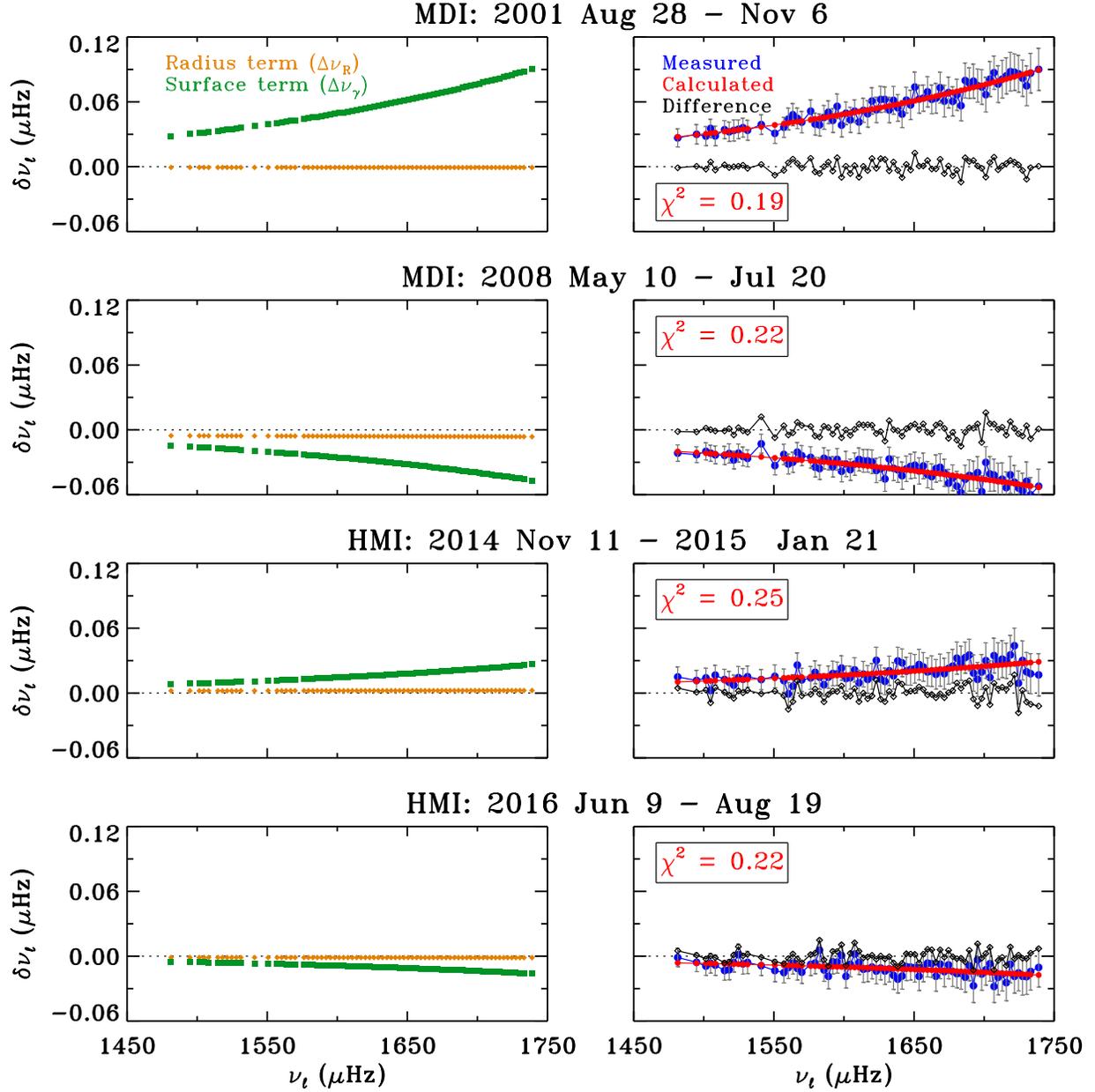}
}
            \caption{(Left) Contributions from two terms on the right-hand side of Equation~\ref{eq1} 
            to {\it f}-mode frequency shifts, and (Right) measured and calculated frequency shifts,
            and the difference between them for four epochs. Positive/negative frequency shifts indicate
            that the reference values are lower/higher than the epochs' values. Errors shown here 
            for the measured shifts.
}
   \label{fitting}
   \end{figure}

 It is clear from Figure~\ref{radius}(a -- b), both $\delta\nu/\nu$ and $\Delta\gamma$ 
 vary in phase with the solar activity while $\Delta R$, as shown in  Figure~\ref{radius}(c), 
 is in anti-phase. Although the solar irradiance increases with decreasing seismic radius 
 or the seismic radius  shrinks with increasing magnetic activity, the anti-correlation between
 them is moderately weak. 
  The shrinkage is believed to be caused by an increase in the 
 radial components of small scale-magnetic field  located a few megameters below the surface
 while  $\Delta\gamma$ depends on a variety of contributions from the near-surface layers. We checked the strength 
 of this correlation/anti-correlation by calculating the Pearson's linear correlation coefficient, 
 $r_P$; 0.99 between $\delta\nu/\nu$ and $F_{10}$,  0.99
 between $\Delta\gamma$ and $F_{10.7}$, and  $-$0.55 between  $\Delta R$ and  TSI. 
 These results are in qualitative agreement with the previous studies \citep{Dziembowski01, Antia04}.
 It should be noted that previous studies were based on the MDI frequencies for first few
 years of cycle 23 only  with a strong 1-year periodicity. 
The contributions from $\Delta\nu_{R}$ and $\Delta\nu_\gamma$ 
 to $\Delta\nu_{\ell}$    are
plotted in Figure~\ref{fitting} (left) for selected epochs; two for MDI and two for HMI.
 It is evident that  $\Delta\nu_{\ell}$ largely depends on $\Delta\nu_\gamma$ while 
 $\Delta\nu_{R}$ has a little contribution.  In right panels, we show the variation of 
 measured and calculated values of $\Delta\nu_{\ell}$, the difference between them. In all cases, we see 
 an increase in $\Delta\nu_\gamma$ with $\ell$, which also increases $\Delta\nu_{\ell}$ values.

In order to quantify the relation between  $\Delta R$ and TSI, we display
the change in TSI ($\Delta$TSI) from its minimum value as a function of corresponding 
change in the seismic radius in Figure~\ref{radius_tsi}. The anti-correlation between
these  two quantities is clearly visible, however the scatter is significantly large 
 for  cycle 23 while  there is a systematic trend in cycle 24. 
These findings are confirmed by calculating $r_P$, which increases from $-$0.51 
for cycle 23 to $-$0.83 for cycle 24. This indicates
that the relationship between  $\Delta R$ and $\Delta$TSI  was weak in cycle 23. 
Note that the solar activity in cycle 24 is significantly reduced as compared to cycle 23 
 and the TSI increased by 0.075\%  from the activity minimum to the maximum. The robustness 
 of the TSI variation per unit change in seismic radius is checked by fitting a
 straight line. The best-fit line is obtained by minimizing the chi-square error statistics.
The $\chi^2$ value for cycle 24 (0.79) is  significantly smaller than that for  cycle 23
(8.17) while for all data sets the  $\chi^2$ is 9.71. This again suggests that $\Delta R$ and $\Delta$TSI had
a stable relation in cycle 24. Further for the 0.1\% TSI increase from
minimum to maximum in an average solar cycle,   the corresponding change in the  seismic radius is about 10 km.
This is 
estimated from the best fit values for cycle 24. The reason for poor correspondence
 between  $\Delta R$ and $\Delta$TSI  in cycle 23 is not clear. 

\begin{figure}   
   \centerline{
\includegraphics[angle=90,scale=0.7]{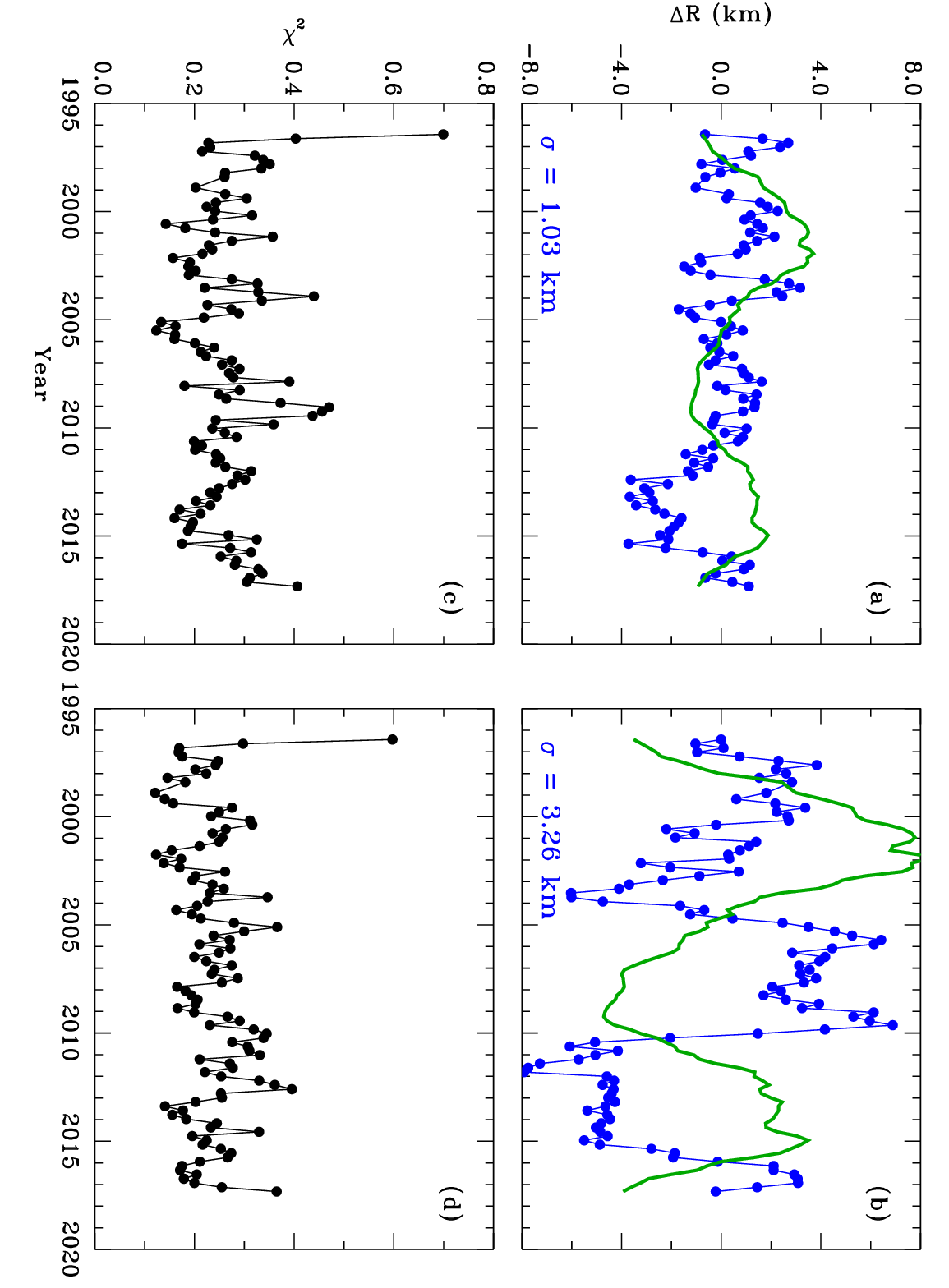}\\
}
            \caption{(Top row) Symbols represent the temporal variation  of  $\Delta R$
             calculated using {\it f}-mode frequencies in (a) low- and
             (b) high-frequency ranges, and  (Bottom row) corresponding $\chi^2$  per degree 
             of freedom from Equation~\ref{eq1}. Solid  green lines in top panels are for
             smoothed TSI values. 
             }
   \label{radius_tsi_nudep}
   \end{figure}

It was suggested by \citet{Sofia05} that the seismic radius change  in shallower layer should 
be larger than that in the deeper layer. Note that we are here analyzing depths within a
few megameters below the photosphere.  In order to  verify the argument by Sofia et al., 
we repeated the analysis 
for {\it f}-modes in two different frequency ranges. The modes in low-$\nu$ range 
(1480 $\mu$Hz $\le \nu  <$ 1630 $\mu$Hz) travel relatively deeper than the modes in 
high-$\nu$ range (1630 $\mu$Hz $\le \nu  <$ 1740 $\mu$Hz). To maintain consistency
we ensured that both frequency ranges cover same number of modes; i.e., 36.
The temporal variations of   $\Delta R$ in these 
frequency ranges are displayed in Figure~\ref{radius_tsi_nudep}.
 One can easily visualize that $\Delta$R in Figure~\ref{radius_tsi_nudep}(b) is larger 
than in Figure~\ref{radius_tsi_nudep}(a).  For low-$\nu$ range,
the correlation coefficients are $-0.09$ and  $-0.80$ corresponding to cycles 23 and 24, respectively, and
these values changed to $-0.48$ and  $-0.75$ for high-$\nu$ range. This again suggests that the TSI
variability had a consistent trend with seismic radius in cycle 24. The standard deviation, $\sigma$, 
for low- and high-$\nu$ ranges are 1.03 km and 3.26 km, respectively and the maximum variation in both cases is
about 2.5$\sigma$. The $\chi^2$ values plotted in lower panels of  Figure~\ref{radius_tsi_nudep} further
indicate that the goodness of fit is not very different in both cases. 

In order to calculate radius change very close
to the surface, one has to use very high-$\ell$ modes, however the helioseismic analysis using
global modes, at present, has  limitations for such studies. It needs to be 
improved to characterize mode parameters  at high degrees because the
ridges  in $\ell$ -- $\nu$ diagrams are not well separated and mode widths are also large  \citep{Korzennik13}.
There are ongoing efforts on the ridge-fitting approach over the traditional mode-fitting method and we hope
 to extend this analysis to higher degrees with different mode sets.
It must be emphasized that the change in seismic radius may be caused by the variation of sound speed, temperature or 
the changes in the superadiabatic superficial layers.  \citet{Dziembowski05} have argued that the decrease
in turbulent pressure or temperature with increasing magnetic activity, or both effects may cause shrinking.  
Furthermore, based on a model of variability of the solar interior with all
observational constraints, \citet{Sofia05} had suggested an increase in seismic  radius variation by
a factor of approximately 1000 from a depth of 5 Mm to the surface, however our study based on
measured frequencies do not support this. In addition, direct measurements of the solar radius 
at the surface also contradict this argument.
 
\section{Summary}

\begin{figure}   
   \centerline{
\includegraphics[scale=0.9]{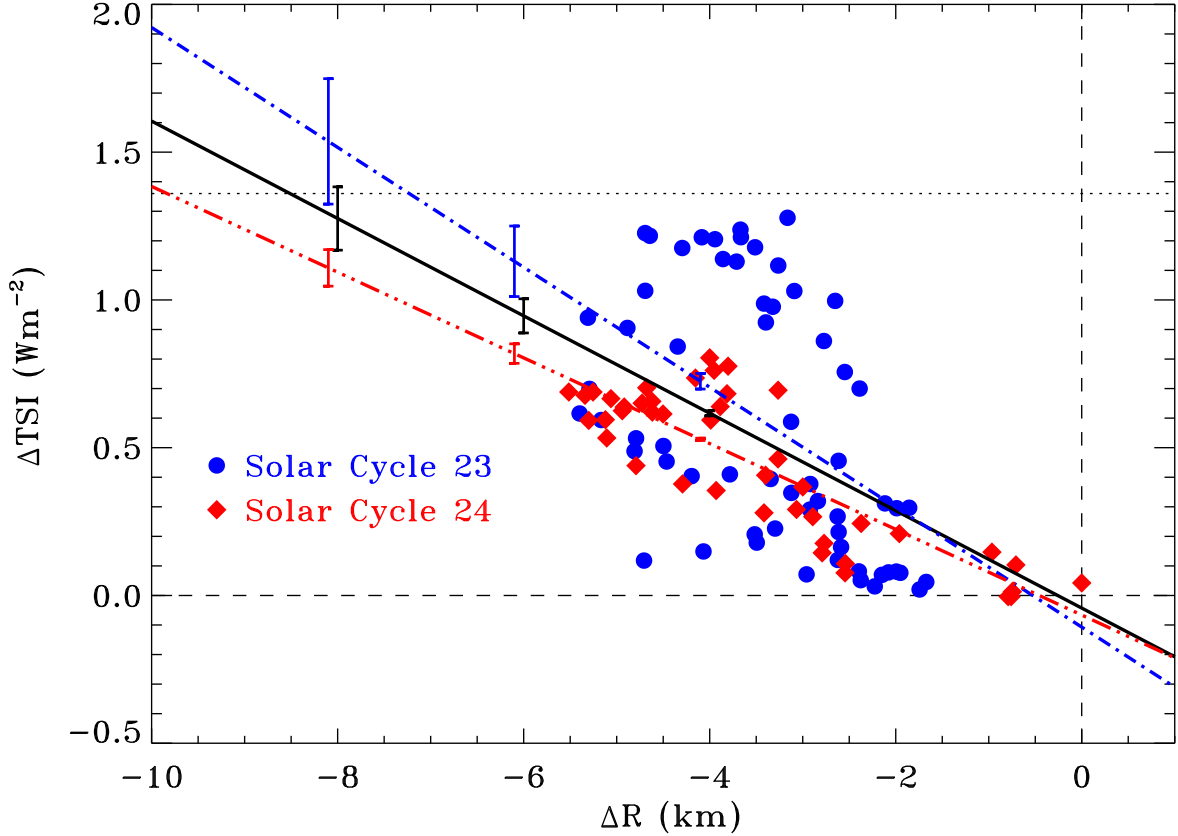}
}
            \caption{Scatter plot showing the variation in TSI with the estimated change in seismic radius in cycle 23
            and cycle 24. 
            Plotted $\Delta R$ and $\Delta TSI$ are the changes 
            from their minimum and maximum values in the entire series, respectively.
            Solid line represents the best linear fit to all data while dashed-dot  and dashed-dot-dot-dot 
            lines are for cycle 23 and 24, respectively. Dotted horizontal line  depicts the 0.1\% change in TSI values from
            the minimum. The errors  shown here are the uncertainties in fitting the straight lines.
}
   \label{radius_tsi}
   \end{figure}

Based on the analysis of  global {\it f}-mode frequencies from MDI and HMI, we demonstrate that 
 the solar irradiance increases with decreasing seismic radius or the seismic
 radius decreases with increasing magnetic activity, however
 these quantities are weakly anti-correlated. We show evidence that 
 the  major contribution to the  TSI variation
comes from the changes in magnetic field while the radius variation plays a secondary role. 
 It must be noted that our results provide information on the changes in thermal structure of 
the outer convection zone only. Although this approach does not provide a precise value of the solar 
radius at the surface,  it is a powerful diagnostic tool to infer changes in the seismic 
radius at a few megameter below the surface. This can also be useful in estimating  changes in
turbulent pressure or temperature with the changing magnetic activity. Based on Figure~\ref{radius_tsi}, 
we estimate that the seismic radius changed by approximately five kilometers during
solar cycles 23 and 24.  The change in seismic radius 
obtained in this study is much smaller than the previous studies where authors have
suggested a radius change of the order of 45 km to explain the 0.1\% variation in
the TSI values \citep{Sofia98}.  

\acknowledgments

 We thank the referee for his comments that have significantly improved the paper.
 We also thank Sarbani Basu for providing useful insight in this work and the
 standard solar model values. This work utilizes data from the {\it SoHO}/MDI and  
{\it SDO}/HMI. {\it SoHO} is a mission of international cooperation
 between ESA and NASA. {\it SDO} data courtesy of {\it SDO} (NASA) and the HMI and AIA consortium. 
The MDI and HMI data are obtained from the Joint Science Operations Center at Stanford University.
The unpublished solar irradiance data set (version v6\_001\_0804) was obtained from VIRGO Team through PMOD/WRC, Davos,
Switzerland.


\clearpage

\clearpage

\clearpage

\clearpage

\clearpage

\clearpage


\begin{thebibliography}{}
\expandafter\ifx\csname natexlab\endcsname\relax\def\natexlab#1{#1}\fi

\bibitem[{{Antia}(1998)}]{Antia98}
{Antia}, H.~M. 1998, \aap, 330, 336

\bibitem[{{Antia} \& {Basu}(2004)}]{Antia04}
{Antia}, H.~M., \& {Basu}, S. 2004, in ESA Special Publication, Vol. 559, SOHO
  14 Helio- and Asteroseismology: Towards a Golden Future, ed. D.~{Danesy}, 301

\bibitem[{{Antia} {et~al.}(2000){Antia}, {Basu}, {Pintar}, \& {Pohl}}]{Antia00}
{Antia}, H.~M., {Basu}, S., {Pintar}, J., \& {Pohl}, B. 2000, \solphys, 192,
  459

\bibitem[{{Bahcall} {et~al.}(2001){Bahcall}, {Pinsonneault}, \&
  {Basu}}]{Bahcall01}
{Bahcall}, J.~N., {Pinsonneault}, M.~H., \& {Basu}, S. 2001, \apj, 555, 990

\bibitem[{{Bahcall} \& {Serenelli}(2005)}]{BS05}
{Bahcall}, J.~N., \& {Serenelli}, A.~M. 2005, \apj, 626, 530

\bibitem[{{Brown} \& {Christensen-Dalsgaard}(1998)}]{Brown98}
{Brown}, T.~M., \& {Christensen-Dalsgaard}, J. 1998, \apjl, 500, L195

\bibitem[{{Dziembowski} \& {Goode}(2005)}]{Dziembowski05}
{Dziembowski}, W.~A., \& {Goode}, P.~R. 2005, \apj, 625, 548

\bibitem[{{Dziembowski} {et~al.}(2001){Dziembowski}, {Goode}, \&
  {Schou}}]{Dziembowski01}
{Dziembowski}, W.~A., {Goode}, P.~R., \& {Schou}, J. 2001, \apj, 553, 897

\bibitem[{{Emilio} {et~al.}(2015){Emilio}, {Couvidat}, {Bush}, {Kuhn}, \&
  {Scholl}}]{Emilio15}
{Emilio}, M., {Couvidat}, S., {Bush}, R.~I., {Kuhn}, J.~R., \& {Scholl}, I.~F.
  2015, \apj, 798, 48

\bibitem[{{Fr{\"o}hlich} \& {Eddy}(1984)}]{Frohlich84}
{Fr{\"o}hlich}, C., \& {Eddy}, J.~A. 1984, Advances in Space Research, 4, 121

\bibitem[{{Fr{\"o}hlich} {et~al.}(1997){Fr{\"o}hlich}, {Crommelynck}, {Wehrli},
  {Anklin}, {Dewitte}, {Fichot}, {Finsterle}, {Jim{\'e}nez}, {Chevalier}, \&
  {Roth}}]{Frohlich97}
{Fr{\"o}hlich}, C., {Crommelynck}, D.~A., {Wehrli}, C., {et~al.} 1997,
  \solphys, 175, 267

\bibitem[{{Gonz{\'a}lez Hern{\'a}ndez} {et~al.}(2009){Gonz{\'a}lez
  Hern{\'a}ndez}, {Scherrer}, \& {Hill}}]{Irene09}
{Gonz{\'a}lez Hern{\'a}ndez}, I., {Scherrer}, P., \& {Hill}, F. 2009, \apjl,
  691, L87

\bibitem[{{Jain} \& {Bhatnagar}(2003)}]{Jain03}
{Jain}, K., \& {Bhatnagar}, A. 2003, \solphys, 213, 257

\bibitem[{{Jain} {et~al.}(2006){Jain}, {Hill}, {Gonz{\'a}lez Hern{\'a}ndez},
  {Toner}, {Tripathy}, {Armstrong}, \& {Jefferies}}]{Jain06a}
{Jain}, K., {Hill}, F., {Gonz{\'a}lez Hern{\'a}ndez}, I., {et~al.} 2006, in ESA
  Special Publication, Vol. 624, Proceedings of SOHO 18/GONG 2006/HELAS I,
  Beyond the spherical Sun, 127.1

\bibitem[{{Kholikov} \& {Hill}(2008)}]{Kholikov08}
{Kholikov}, S., \& {Hill}, F. 2008, \solphys, 251, 157

\bibitem[{{Korzennik} {et~al.}(2013){Korzennik}, {Rabello-Soares}, {Schou}, \&
  {Larson}}]{Korzennik13}
{Korzennik}, S.~G., {Rabello-Soares}, M.~C., {Schou}, J., \& {Larson}, T.~P.
  2013, \apj, 772, 87

\bibitem[{{Larson} \& {Schou}(2015)}]{Larson15}
{Larson}, T.~P., \& {Schou}, J. 2015, \solphys, 290, 3221

\bibitem[{{Larson} \& {Schou}(2018)}]{Larson18}
---. 2018, \solphys, 293, 29

\bibitem[{{Menezes} \& {Valio}(2017)}]{Menezes17}
{Menezes}, F., \& {Valio}, A. 2017, ArXiv e-prints, arXiv:1712.06771

\bibitem[{{Pap} {et~al.}(2001){Pap}, {Rozelot}, {Godier}, \& {Varadi}}]{Pap01}
{Pap}, J., {Rozelot}, J.~P., {Godier}, S., \& {Varadi}, F. 2001, \aap, 372,
  1005

\bibitem[{{Rozelot} {et~al.}(2015){Rozelot}, {Kosovichev}, \&
  {Kilcik}}]{Rozelot15}
{Rozelot}, J.~P., {Kosovichev}, A., \& {Kilcik}, A. 2015, \apj, 812, 91

\bibitem[{{Scherrer} {et~al.}(1995){Scherrer}, {Bogart}, {Bush}, {Hoeksema},
  {Kosovichev}, {Schou}, {Rosenberg}, {Springer}, {Tarbell}, {Title},
  {Wolfson}, {Zayer}, \& {MDI Engineering Team}}]{mdi}
{Scherrer}, P.~H., {Bogart}, R.~S., {Bush}, R.~I., {et~al.} 1995, \solphys,
  162, 129

\bibitem[{{Scherrer} {et~al.}(2012){Scherrer}, {Schou}, {Bush}, {Kosovichev},
  {Bogart}, {Hoeksema}, {Liu}, {Duvall}, {Zhao}, {Title}, {Schrijver},
  {Tarbell}, \& {Tomczyk}}]{hmi}
{Scherrer}, P.~H., {Schou}, J., {Bush}, R.~I., {et~al.} 2012, \solphys, 275,
  207

\bibitem[{{Schou} {et~al.}(1997){Schou}, {Kosovichev}, {Goode}, \&
  {Dziembowski}}]{Schou97}
{Schou}, J., {Kosovichev}, A.~G., {Goode}, P.~R., \& {Dziembowski}, W.~A. 1997,
  \apjl, 489, L197

\bibitem[{{Schou} {et~al.}(2002){Schou}, {Howe}, {Basu},
  {Christensen-Dalsgaard}, {Corbard}, {Hill}, {Komm}, {Larsen},
  {Rabello-Soares}, \& {Thompson}}]{Schou02}
{Schou}, J., {Howe}, R., {Basu}, S., {et~al.} 2002, \apj, 567, 1234

\bibitem[{{Sofia}(1998)}]{Sofia98}
{Sofia}, S. 1998, in Solar Electromagnetic Radiation Study for Solar Cycle 22,
  ed. J.~M. {Pap}, C.~{Frohlich}, \& R.~K. {Ulrich}, 413

\bibitem[{{Sofia} {et~al.}(2005){Sofia}, {Basu}, {Demarque}, {Li}, \&
  {Thuillier}}]{Sofia05}
{Sofia}, S., {Basu}, S., {Demarque}, P., {Li}, L., \& {Thuillier}, G. 2005,
  \apjl, 632, L147

\bibitem[{{Sofia} {et~al.}(1979){Sofia}, {O'keefe}, {Lesh}, \&
  {Endal}}]{Sofia79}
{Sofia}, S., {O'keefe}, J., {Lesh}, J.~R., \& {Endal}, A.~S. 1979, Science,
  204, 1306

\bibitem[{{Tapping}(2013)}]{Tapping13}
{Tapping}, K.~F. 2013, Space Weather, 11, 394

\bibitem[{{Thuillier} {et~al.}(2017){Thuillier}, {Zhu}, {Shapiro}, {Sofia},
  {Tagirov}, {van Ruymbeke}, {Perrin}, {Sukhodolov}, \&
  {Schmutz}}]{Thuillier17}
{Thuillier}, G., {Zhu}, P., {Shapiro}, A.~I., {et~al.} 2017, \aap, 603, A28

\bibitem[{{Tripathy} \& {Antia}(1999)}]{Tripathy99}
{Tripathy}, S.~C., \& {Antia}, H.~M. 1999, \solphys, 186, 1

\bibitem[{{Vaquero} {et~al.}(2016){Vaquero}, {Gallego}, {Ruiz-Lorenzo},
  {L{\'o}pez-Moratalla}, {Carrasco}, {Aparicio}, {Gonz{\'a}lez-Gonz{\'a}lez},
  \& {Hern{\'a}ndez-Garc{\'{\i}}a}}]{Vaquero16}
{Vaquero}, J.~M., {Gallego}, M.~C., {Ruiz-Lorenzo}, J.~J., {et~al.} 2016,
  \solphys, 291, 1599

\end{thebibliography}

\end{document}